\documentclass[a4paper,fleqn,usenatbib]{mnras}

\usepackage[T1]{fontenc}
\usepackage{ae,aecompl}

\usepackage{graphicx}	
\usepackage{amsmath}	
\usepackage{amssymb}	
\usepackage{multirow}
\usepackage{booktabs}

\newcommand{\integral}{\textit{INTEGRAL}}

\newcommand{\nustar}{\textit{NuSTAR}}
\newcommand{\suzaku}{{\it Suzaku}}
\newcommand{\swift}{{\it Swift}}
\newcommand{\xmm}{{\it XMM-Newton}}

\newcommand{\igr}{IGR~2124.7+5058}
\newcommand{\eso}{ESO~103--035}

\title[Coronal temperatures with \nustar]{Coronal temperatures of the AGN \eso\ and \igr\ from \nustar\ observations}

\author[D. J. K. Buisson et al.]{
	D. J. K. Buisson$^{1}$,\thanks{Email: djkb2@ast.cam.ac.uk}
	A. C. Fabian$^{1}$ and
	A. M. Lohfink${^2}$\\
  $^{1}$Institute of Astronomy, Madingley Road, Cambridge, CB3 0HA\\
  $^{2}$Department of Physics, Montana State University, Bozeman, 59717-3840, MT, USA
  }

\date{Accepted 2018 September 19. Received 2018 September 17; in original form 2018 August 29}

\pubyear{2018}

\begin{document}
\label{firstpage}
\pagerange{\pageref{firstpage}--\pageref{lastpage}}
\maketitle

\begin{abstract}
We present measurements of the coronae of two AGN from hard X-ray observations made with \nustar:
\eso, a moderately to highly obscured source with significant reflection; and \igr, a radio-loud source with a very hard spectrum.
Using an exponentially cut-off powerlaw model for the coronal emission spectrum gives a high-energy cut-off of $100_{-30}^{+90}$\,keV for \eso\ and $80_{-9}^{+11}$\,keV for \igr, within the typical range for AGN. Fitting with physical Comptonisation models shows that these correspond to a temperature of $22_{-6}^{+19}$ and $20_{-2}^{+3}$\,keV respectively.
These values are consistent with pair production limiting the coronal temperature.
\end{abstract}

\begin{table*}
\caption{List of observations of \eso\ and \igr. Exposure is the mean good exposure per FPM, as used for spectral fitting.}
\label{tab:obs}
\centering

\begin{tabular}{llclrlr}
\hline
\multirow{2}{*}{Source} &\multirow{2}{*}{Campaign} & \multicolumn{3}{c}{\nustar} & \multicolumn{2}{c}{\swift}\\
\cmidrule(lr){3-5}\cmidrule(lr){6-7}
 &  & OBSID & Start date & Exposure/ks & \swift\ OBSIDs & Exposure/ks\\
\hline
\multirow{2}{*}{\eso} & EGS & 60061288002 & 2013-02-25 & 27.3 & 00080219001 & 6.7 \\
& Cycle~3 & 60301004002 & 2017-10-15 & 42.5 & 00088112001 & 1.9 \\ 
\hline
\multirow{2}{*}{\igr} & EGS & 60061305002 & 2014-12-13 & 23.9 & 00080273001/2 & 6.8 \\
& Cycle~3 & 60301005002 & 2018-01-02 & 40.2 & 00088113001/2/3 & 4.0\\
\hline
\end{tabular}
\end{table*}

\section{Introduction}

Active Galactic Nuclei (AGN) are powered by accretion onto a supermassive black hole (SMBH), converting gravitational potential energy to radiation across the electromagnetic spectrum. Due to the shape of the gravitational potential well, the majority of the energy is released in the innermost few gravitational radii ($r_g=GM_{\rm BH}/c^2$). Localised to this region is the X-ray emitting corona, which Compton scatters incident optical and UV photons to X-ray energies \citep[e.g.][]{haardt91} and is typically regarded as a region of electron pair plasma. 

The X-ray spectrum of emission from the corona may be approximated by a powerlaw up to some cut-off energy where emission quickly rolls over \citep{rybicki79}. The index of this powerlaw and the energy at which the cut-off occurs are then the primary observable characteristics from which conditions in the corona may be inferred.

Since the high-energy cut-off occurs when the electrons are no longer able to add energy to the photons in an interaction, its value is governed by the electron temperature (if the particles in the corona have a roughly thermal spectrum).
If the cut-off is modelled as an exponential suppression of the emission ($N(E)\propto E^{-\Gamma}e^{-E/E_{\rm Cut}}$), the value inferred is around 2--3 times the temperature (\citealt{petrucci01}, where energy and temperature are expressed in the same units by $E=k_{\rm B}T$).

The hard X-ray surveys performed by \integral\ \citep{malizia14} and \swift-BAT \citep{vasudevan13,ricci17} have shown that this cut-off is typically around a few hundred keV: \citet{malizia14} find a median of $128$\,keV and a standard devation of $46$\,keV; \citet{ricci17} find a median of $200\pm29$\,keV.
The cut-off energy also seems to decrease with Eddington rate \citep{ricci18}.

The mechanism by which the coronal temperature is regulated is, however, still an open question. One possibility is (electron) pair production in photon-photon collisions, the rate of which increases rapidly above a certain temperature. This provides many more particles to share the energy and so makes further temperature increase difficult \citep{bisnovatyi71,svensson82,guilbert83,svensson84}. This temperature then acts as an effective upper limit for the electron temperature.
This possibility was explored in \citet{fabian15} and found to be reasonable: sources were seen to have temperatures close to the limit imposed by pair production.

Observations from \nustar\ \citep{harrison13} are able to refine this picture: owing to its ability to focus hard (up to 78\,keV) X-rays, \nustar\ allows more precise measurements to be made of dimmer sources with shorter observations. This increased signal also allows the effect of degeneracy between curvature due to the cut-off and due to reflection to be reduced.

Here, we present new studies of the coronae of two AGN, \eso\ and \igr, from recent \nustar\ observations. 

\subsection{ESO~103--035}

ESO~103--035 ($z=0.013$) is an optical Seyfert~2 galaxy \citep{veron06} initially detected in X-rays with HEAO-A2 \citep{marshall79,phillips79}. {\it EXOSAT} observations showed absorption with variability by almost a factor of 2 in column density over 90 days, from $1.7$ to $1.0\times10^{23}$\,cm$^{-2}$ \citep{warwick88}.
ESO~103--035 was also observed with {\it BeppoSAX}, in October of 1996 and 1997 \citep{wilkes01,akylas01}, again finding significant absorption ($N_{\rm H}=1.79\pm0.09\times10^{23}$\,cm$^{-2}$) and also an iron-K emission line. \citet{wilkes01} additionally find an iron absorption edge and a low cut-off ($29\pm10$\,keV).

Furthermore, the galaxy contains a nuclear maser source \citep{bennert04} and the black hole mass has been estimated as $M_{\rm BH}=10^{7.1\pm0.6}\,{\rm M_{\odot}}$ \citep{czerny01}.

The Galactic absorption column is modest, $N_{\rm H, Gal} = 4.56-6.81\times10^{20}$ \citep{kalberla05}, $6.42-7.86\times10^{20}$\,cm$^{-2}$ \citep{dickey90}.

\subsection{IGR~2124.7+5058 (4C~50.55)}

IGR J21247+5058 (4C~50.55, $z=0.02$, \citealt{masetti04}) is a bright radio loud Seyfert~1 galaxy. Optical studies of this source have been challenging due to its alignment with a Galactic star \citep{masetti04}.

Several X-ray missions have observed \igr. \citet{molina07} analyse \xmm\ data, finding significant absorption (up to $10^{23}$\,cm$^{-2}$) and weak reflection. Combining the \xmm\ data with \integral\ data constrains the high-energy cut-off to $E_{\rm Cut}=100_{-30}^{+55}$\,keV.
The addition of \swift-BAT data refines this to $79_{-15}^{+23}$\,keV.

\citet{tazaki10} apply Comptonisation models to \suzaku\ observations, finding $\tau_{\rm e}\sim3$ and $kT_{\rm e}\sim30$\,keV. Their modelling of the Fe K-$\alpha$ line finds an inner disc radius $R_{\rm in}\sim700\,r_{\rm g}$, which they explain by the inner disc being either truncated or covered by the corona. The flux is stable throughout most of the 170\,ks observation but increases by 30\,\% below 10\,keV in the last 20\,keV.

\igr\ is a radio-loud source, so it is possible that the X-ray spectrum includes a contribution from a jet. \citet{tazaki10} calculate the likely contribution based on the radio to gamma-ray SED and conclude that any contribution is between $10^{-4}$ and $10^{-1}$ of the X-ray power in observations of similar flux to those analysed here.

The Galactic absorption to \igr\ is significant, being measured at $N_{\rm H, Gal} = 0.855-1.16\times10^{22}$ \citep{kalberla05}, $1.02-1.39\times10^{22}$\,cm$^{-2}$ \citep{dickey90}.
Since the total absorption to \igr\ is higher still and the redshift is low, differences in Galactic emission are degenerate with intrinsic absorption, so we fix Galactic absorption to $10^{22}$\,cm$^{-2}$.

\section{Observations and Data Reduction}
\label{section_datareduction}

There are two \nustar\ observations of each source, separated by several years; each observation has a simultaneous \swift\ snapshot (see Table~\ref{tab:obs}). For each source, one observation was made as part of the \nustar\ Extragalactic Survey (EGS) and one as a Cycle~3 Guest Observer target; we therefore refer to the observations as `EGS' and `Cycle 3'.

We reduced the \nustar\ data with \textsc{NuSTARDAS} version 1.8.0 and CALDB version 20171002.
We produced clean event files using \textsc{nupipeline}, choosing filtering options for the SAA based on the online background reports. In each case the option which gave the greatest exposure while removing periods of elevated background was \textsc{saacalc=2 saamode=optimized tentacle=yes}.
Spectra and associated response files were produced using the \textsc{nuproducts} command, with a 60\,arcsec radius circular source region and a 90\,arcsec radius circular background region from a source free area of the same chip (the largest such region available).

The \swift-XRT data were reduced using the online \swift-XRT products generator\footnote{www.swift.ac.uk/user\_objects}, as described in \citet{evans09}. We extracted the mean spectrum of the \swift\ observation(s) associated with each \nustar\ observation (see Table~\ref{tab:obs}).

For \eso, the high absorption column means that the XRT data provide little signal below 3\,keV (only one bin with the grouping used) and the greater effective area of \nustar\ means this data dominates above 3\,keV, so we do not use XRT data in spectral fits of \eso.

We also compare with the \swift-BAT data of the sources. We use the spectra from the 105 month catalogue\footnote{swift.gsfc.nasa.gov/results/bs105mon/} \citep{oh18} and light curves from the transient monitor\footnote{swift.gsfc.nasa.gov/results/transients/} \citep{krimm13}.

Spectra from all instruments (apart from \swift-BAT) were grouped to a signal to noise level of 6. Fits were made in ISIS Version 1.6.2-42 \citep{houck00}; errors are given at the 90\% level.
We use the elemental abundances of \citet{wilms00} with cross sections from \citet{verner96}.

\section{Results}
\label{sec:results}

\begin{figure*}
\centering
\includegraphics[width=\columnwidth]{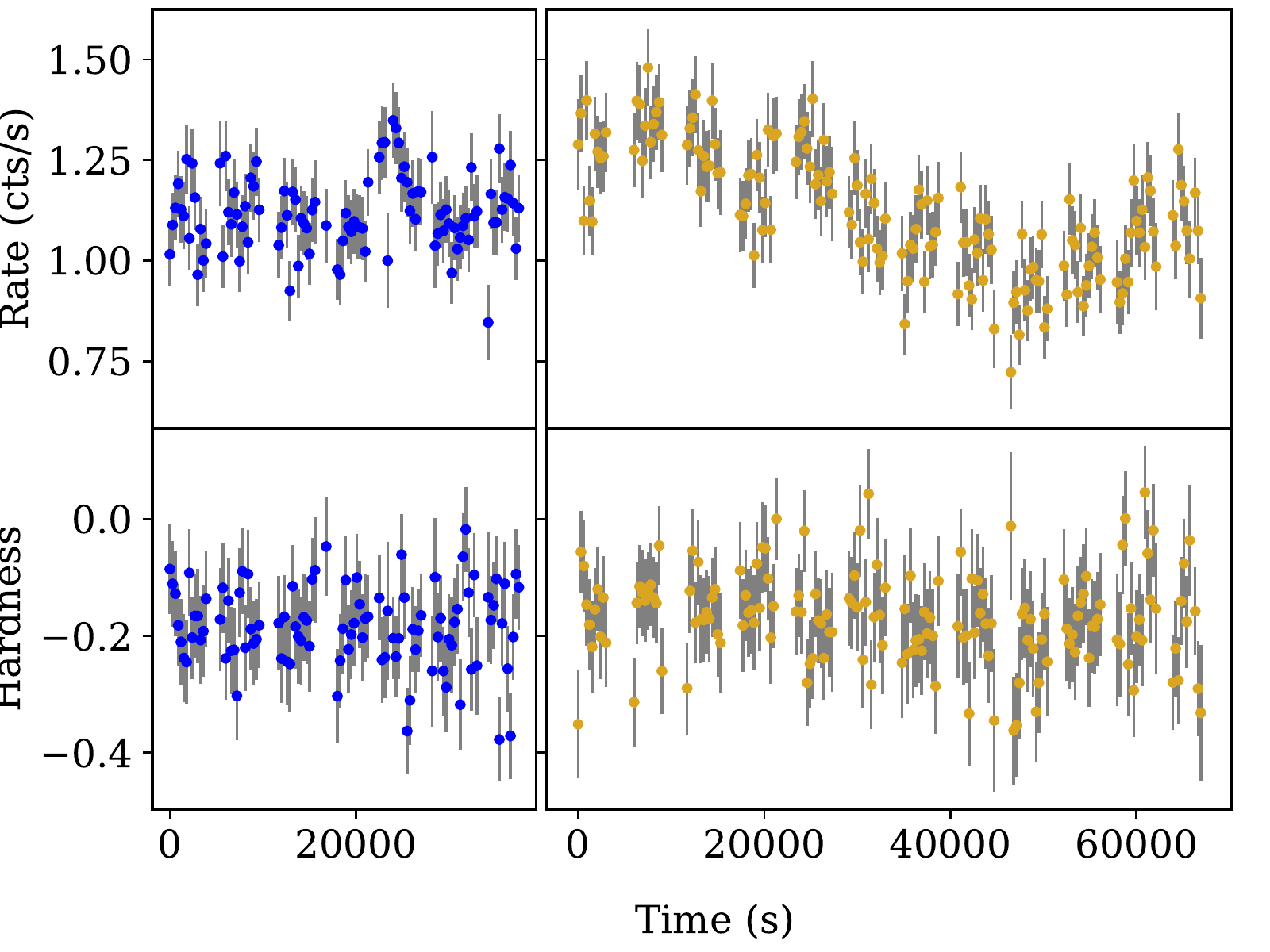}
\includegraphics[width=\columnwidth]{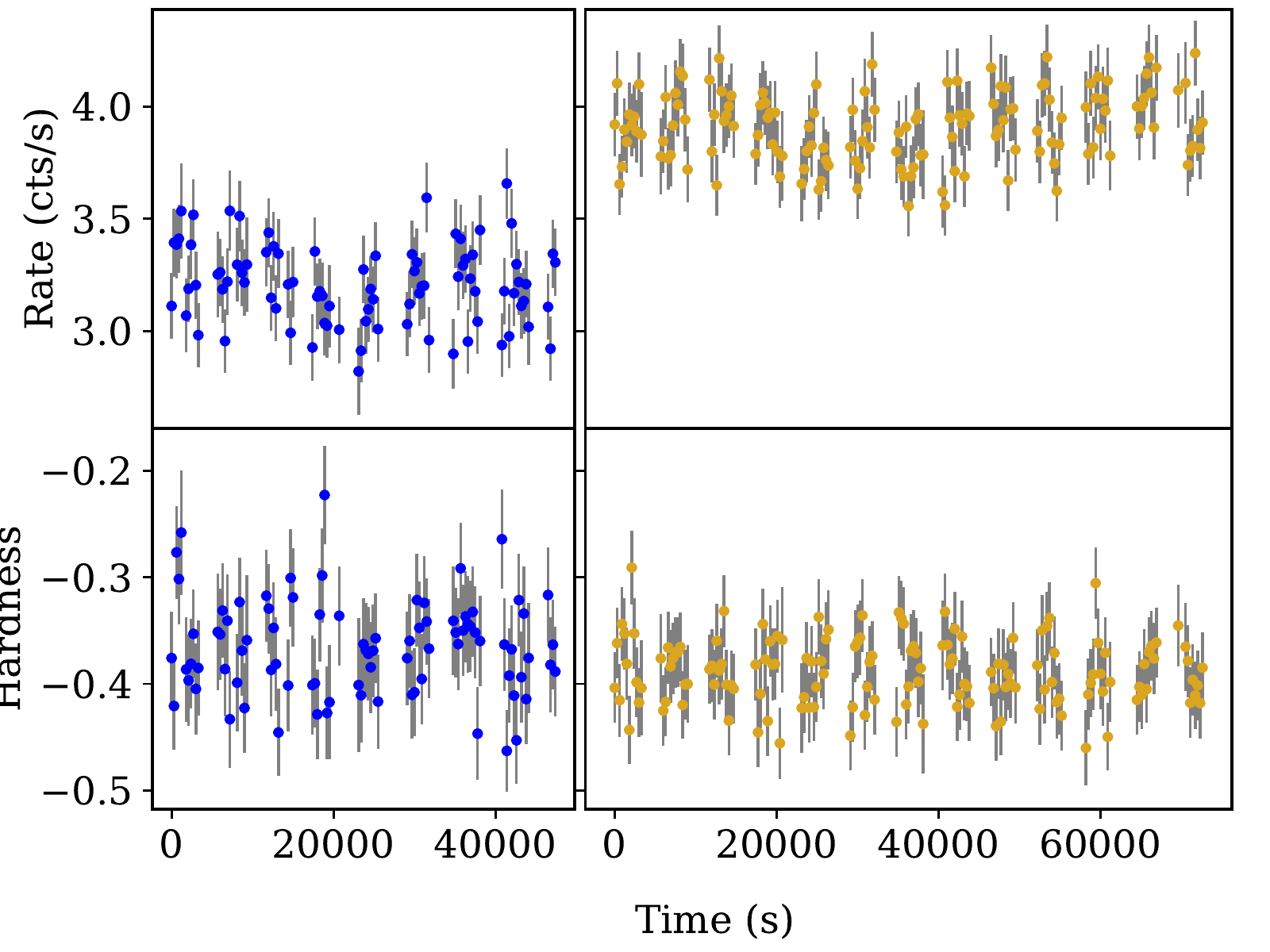}
\caption{\nustar\ (FPMA) light curve and hardness ratio with 300\,s bins for \eso\ (left) and \igr\ (right). The first (blue) curve for each source shows the EGS observation, the second (yellow) cycle 3. The rate is given for $3-78$\,keV. Hardness is defined as $(H-S)/(H+S)$, where $H$ is $10-50$\,keV rate and $S$ is $3-10$\,keV rate. 
}
\label{fig_nustarlightcurve}
\end{figure*}

\begin{figure*}
\centering
\includegraphics[width=\columnwidth]{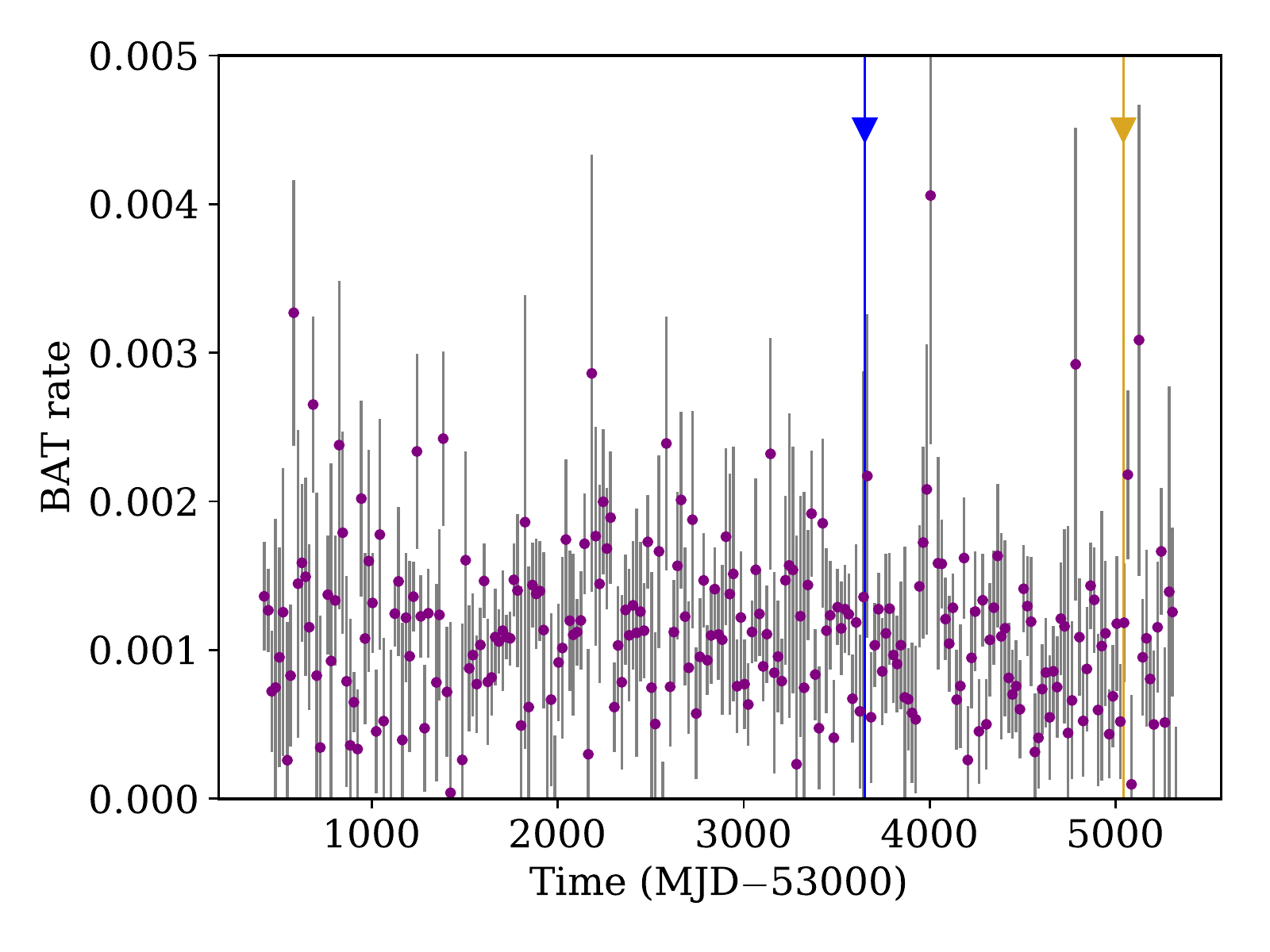}
\includegraphics[width=\columnwidth]{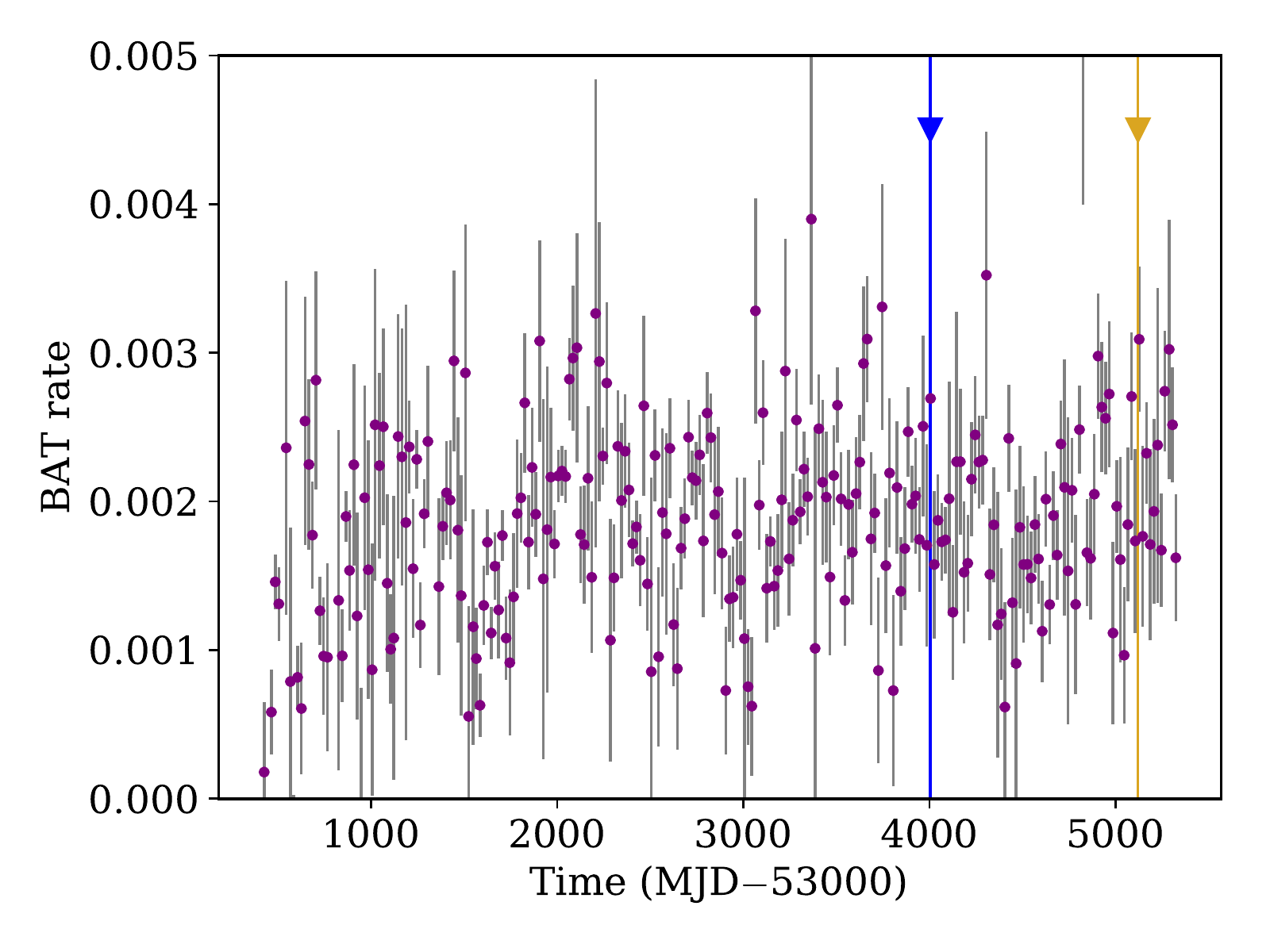}
\caption{\swift-BAT light curves of \eso\ (left) and \igr\ (right), binned to 20\,days, with times of \nustar\ observations shown as vertical lines.
}
\label{fig_batlightcurve}
\end{figure*}

We begin by producing a light curve and hardness ratio for each source (Fig.~\ref{fig_nustarlightcurve}).
While both sources have changed in flux between their two observations, the light curves show little variability within an observation for \igr\ and moderate slow variability for \eso. Additionally, there is little change in hardness within any observation: each  observation is consistent with constant hardness ($\chi^2/{\rm d.o.f}=84/99$ and $147/157$ for \eso; $89/86$ and $121/144$ for \igr). Therefore, we extract mean spectra from the whole of each observation of each source.

\begin{figure*}
\centering
\includegraphics[width=\columnwidth]{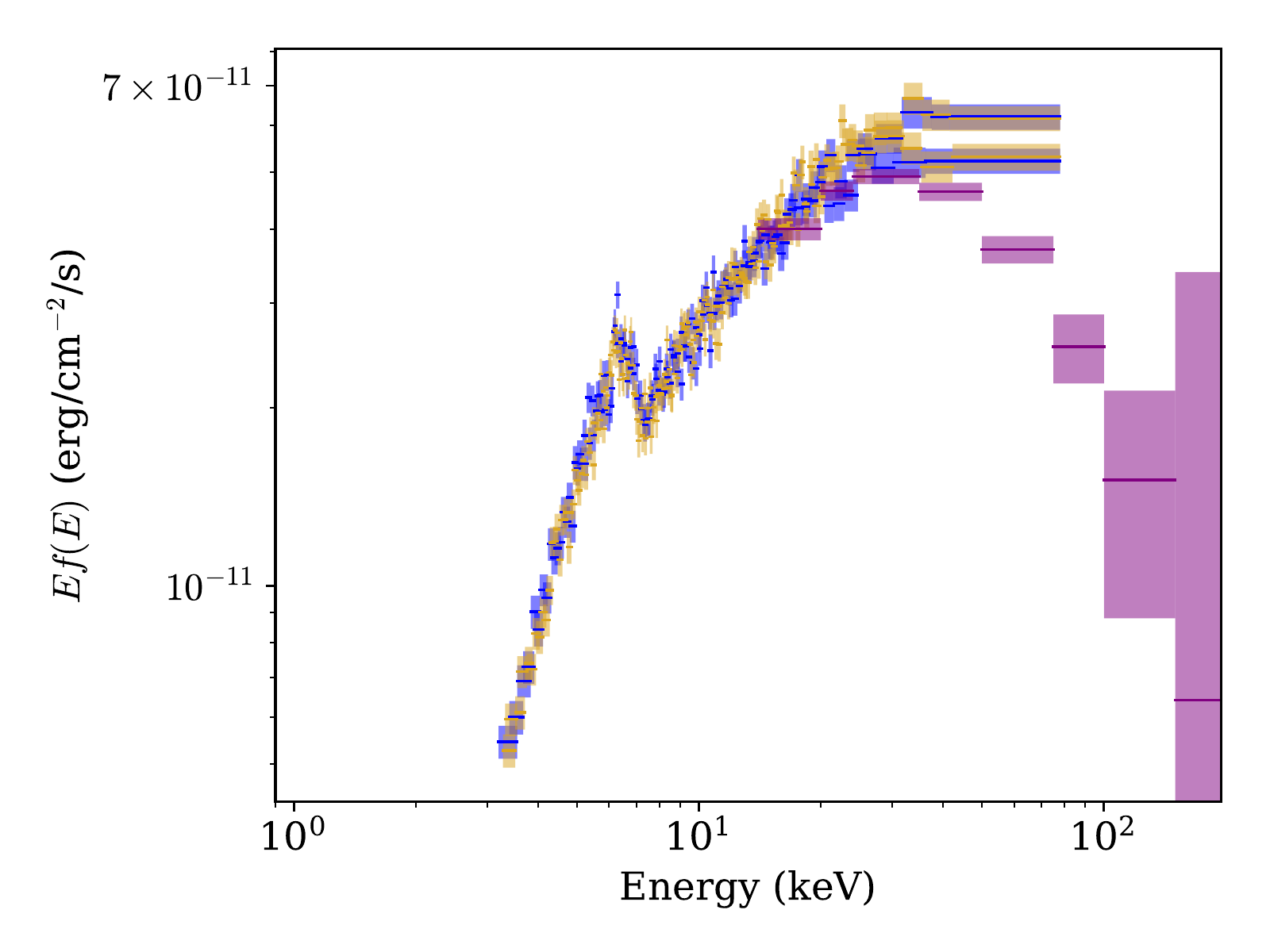}
\includegraphics[width=\columnwidth]{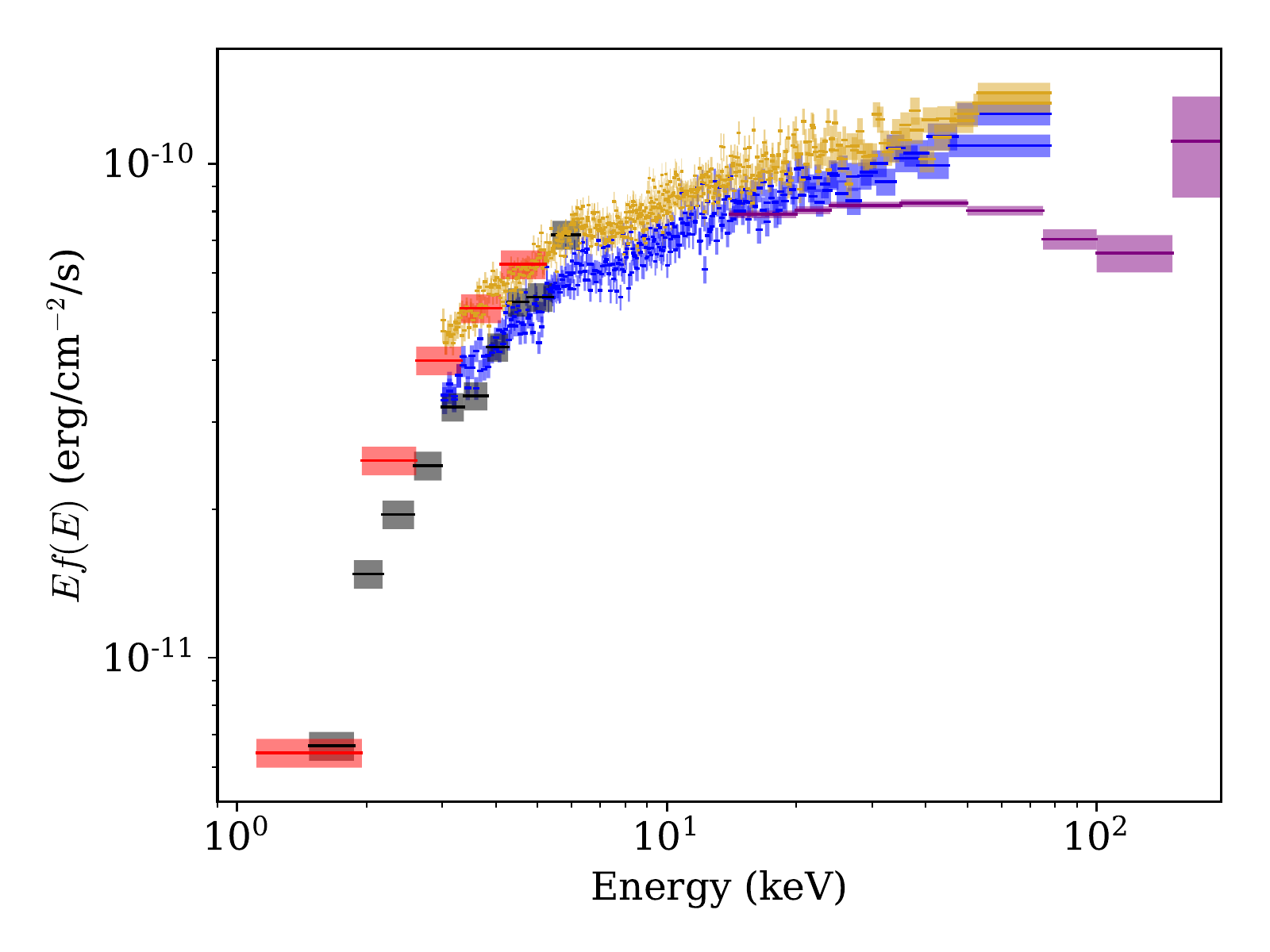}
\caption{Unfolded spectra of \eso\ (left) and \igr\ (right). Both sources have hard, absorbed spectra. \eso\ shows similar hard-energy emission to the long-term average from \swift-BAT; \igr\ is brighter and harder in the \nustar\ observations than the average.
\swift-XRT ($<10$\,keV) is shown in black (EGS) and red (Cycle 3); \nustar\ ($3-78$\,keV) in blue (EGS) and yellow (Cycle 3); and \swift-BAT ($15-200$\,keV) in purple.}
\label{fig:unfold}
\end{figure*}

We show the spectra unfolded against a constant model ($\Gamma=2$ powerlaw) in Fig.~\ref{fig:unfold}.
Each source shows a hard spectrum with significant absorption. \eso\ matches the long-term \swift-BAT spectrum well but \igr\ exceeds the BAT flux by almost a factor of 2 at high energies (within the \nustar\ band). This higher flux is consistent with the variability in the long-term BAT light curve (Figure~\ref{fig_batlightcurve}).

\begin{figure*}
\centering
\includegraphics[width=\columnwidth]{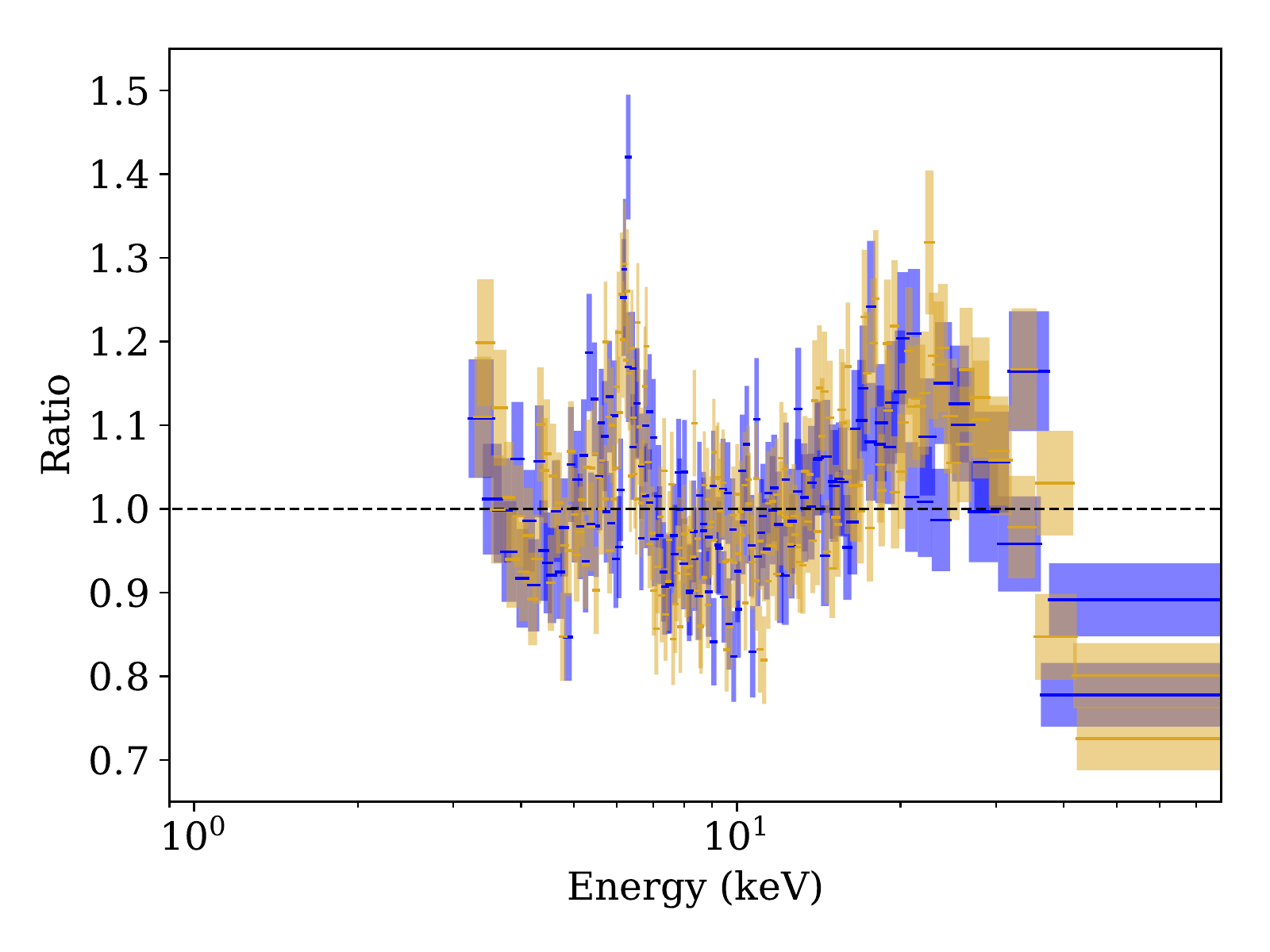}
\includegraphics[width=\columnwidth]{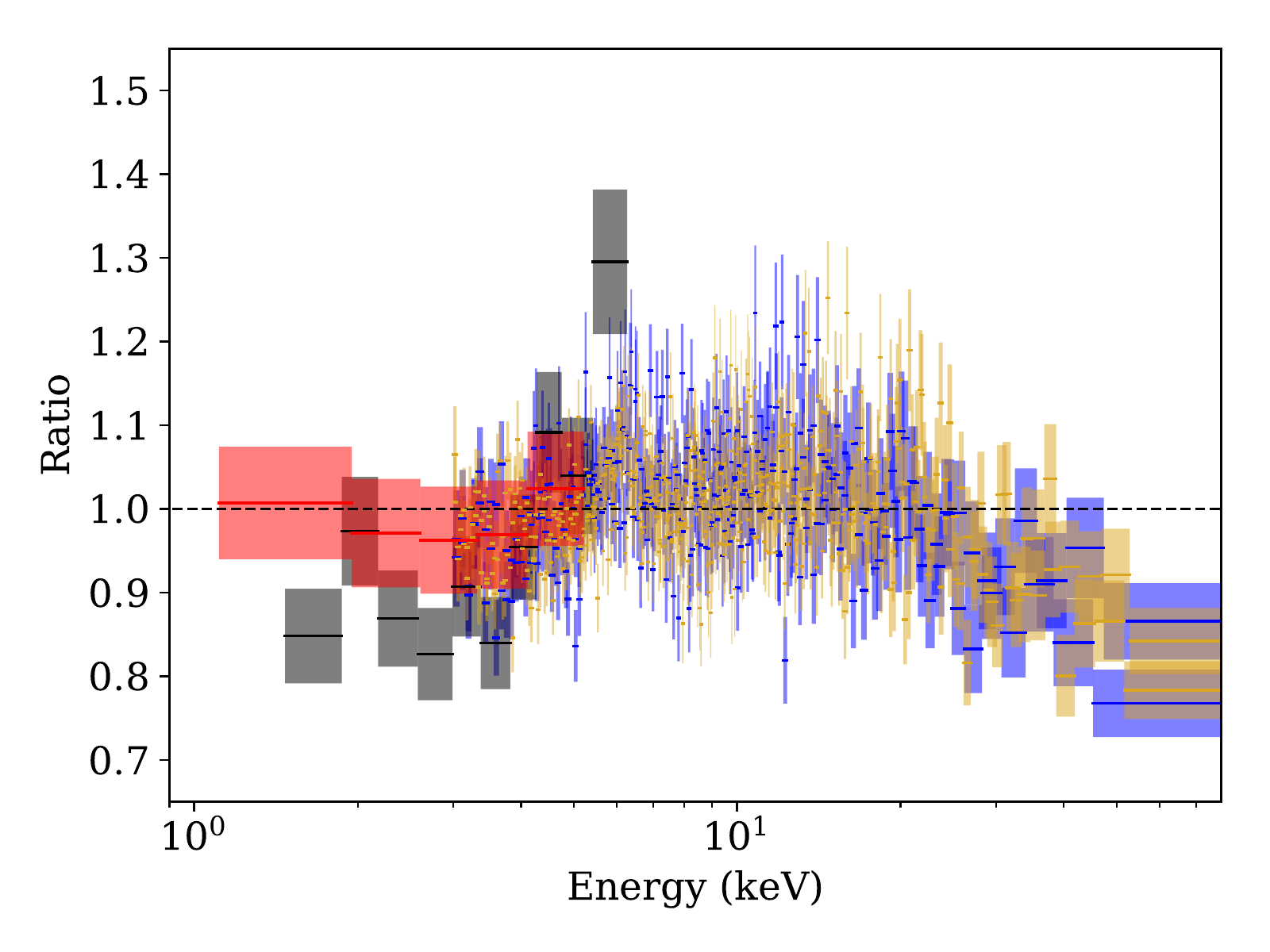}
\caption{Ratio of spectra of \eso\ (left) and \igr\ (right) to an absorbed powerlaw. For each source, the absorption is fixed to the best fit value from fits presented later and powerlaw parameters are fit to each observation separately. Both sources show a roll-over at high energies, while reflection features are stronger in \eso.}
\label{fig:ratio}
\end{figure*}

To show spectral features more clearly, we also plot the ratio of each spectrum to an absorbed powerlaw. Since this ratio is primarily for display, we fix the absorption to match the best-fit from detailed modelling performed later and fit for the power law normalisation and slope.
Both sources show a drop in flux relative to the simple power law at high energies. \eso\ shows a strong iron line and Compton hump indicative of reflected emission, while \igr\ shows these features only more weakly.

\begin{figure*}
\centering
\includegraphics[width=\columnwidth]{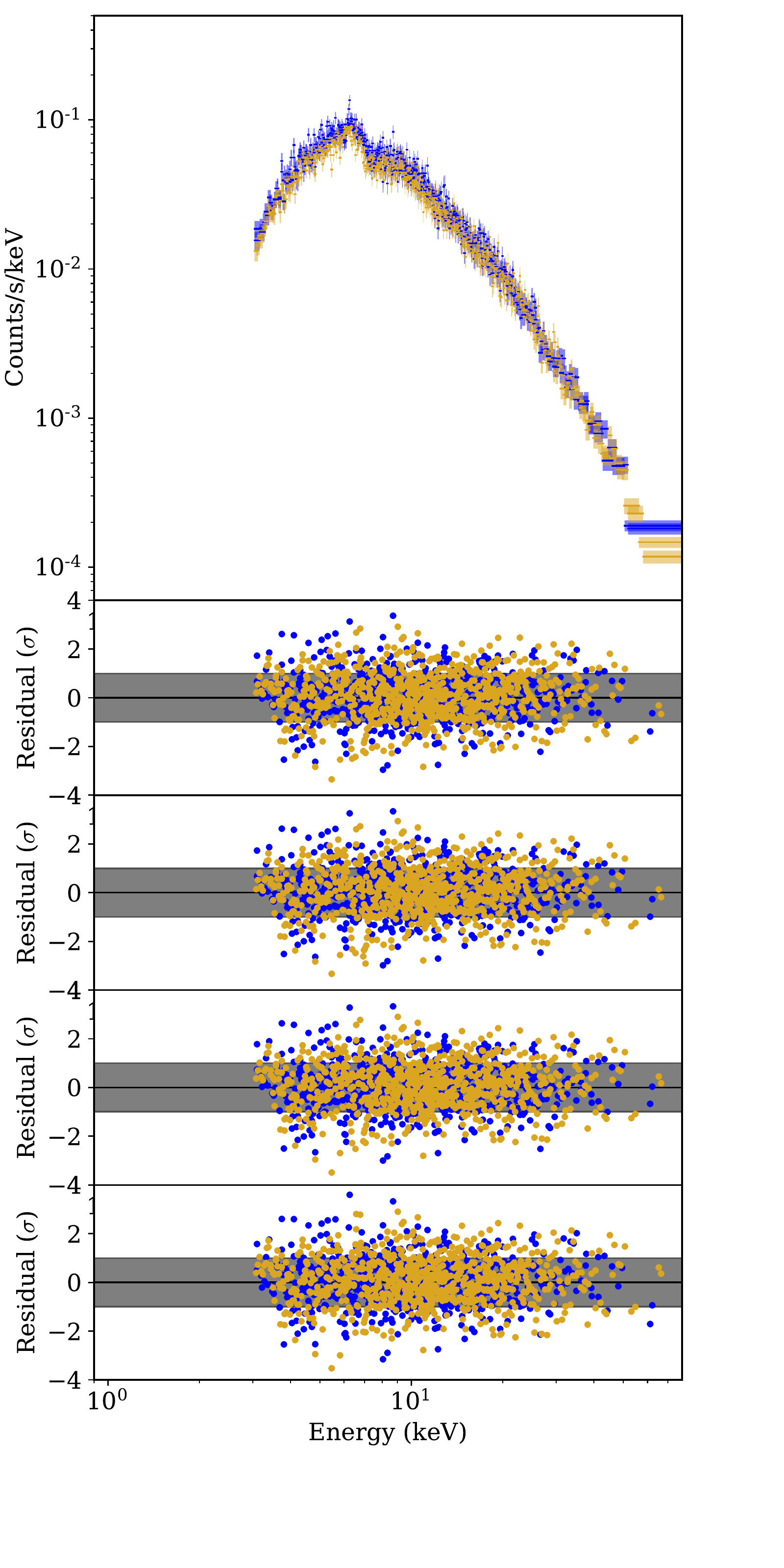}
\includegraphics[width=\columnwidth]{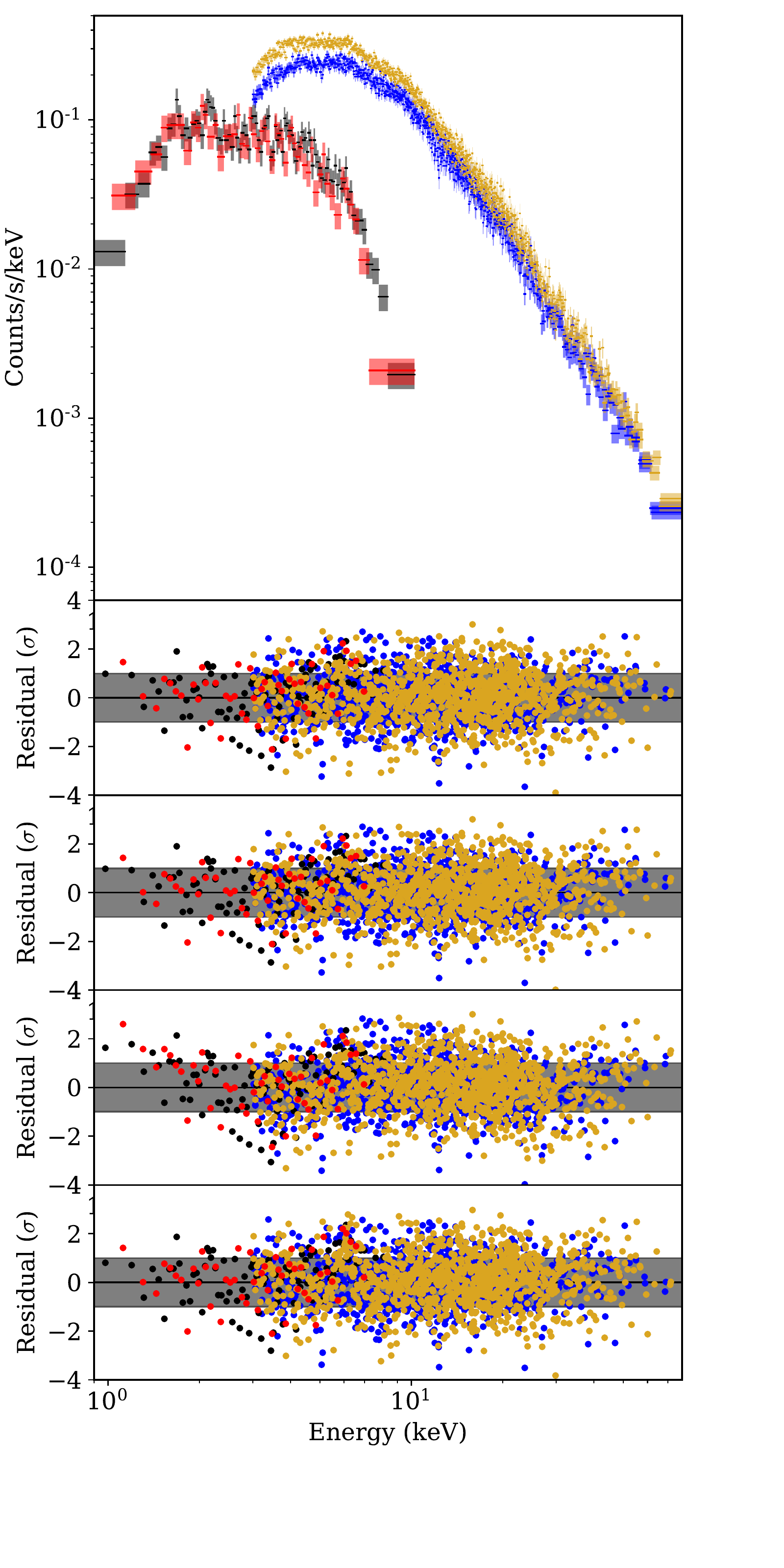}
\caption{Data and residuals of best fitting broadband models for \eso\ (left) and \igr\ (right). Top: data; lower: residuals to models: EGS in blue/black, Cycle 3 in red/yellow. From top to bottom: \textsc{pexmon, xillver, xillverCp, relxillCp}.}
\label{fig:fitres}
\end{figure*}

\begin{table*}
\begin{tabular}{lcccccccc}
\hline
Dataset & \multicolumn{4}{c}{EGS} & \multicolumn{4}{c}{Cycle~3} \\
\cmidrule(lr){2-5}\cmidrule(lr){6-9}
Model & \textsc{pexmon} & \textsc{xillver} & \textsc{xillverCp} & \textsc{relxillCp}& \textsc{pexmon} & \textsc{xillver} & \textsc{xillverCp} & \textsc{relxillCp} \\
\hline
$N_{\rm H}$/$10^{22}$cm$^{-2}$ & $17.1_{-1.4}^{+1.5}$ & $17.0_{-1.6}^{+1.8}$ & $17.3_{-1.3}^{+1.6}$ & $17.0_{-1.6}^{+1.0}$ & $15.6_{-1.2}^{+1.2}$ & $15_{-1.3}^{+1.4}$ & $16.4_{-0.9}^{+1.2}$ & $16_{-1.1}^{+1.0}$\\
$\Gamma$ & $1.82_{-0.16}^{+0.19}$ & $1.79_{-0.16}^{+0.24}$ & $1.84_{-0.10}^{+0.19}$ & $1.86_{-0.16}^{+0.13}$ & $1.71_{-0.13}^{+0.15}$ & $1.73_{-0.15}^{+0.22}$ & $1.82_{-0.08}^{+0.16}$  & $1.76_{-0.08}^{+0.07}$\\
$E_{\rm Cut}$/keV &$130_{-60}^{+450}$  & $110_{-40}^{+340}$ & - & - & $100_{-30}^{+90}$ & $100_{-30}^{+160}$ & - & - \\
$kT_{\rm e}$/keV & - & - &  $>17$ & $>20$ & - & - & $27_{-9}^{+200}$ & $22_{-6}^{+19}$\\
$A_{\rm Fe}$ & $0.8_{-0.3}^{+0.4}$ & $1.1_{-0.5}^{+1}$ & $1.5_{-0.9}^{+1.6}$ & $<7$ & $0.8_{-0.2}^{+0.3}$ & $0.9_{-0.3}^{+0.5}$ & $1.0_{-0.4}^{+0.8}$ & $2.0_{-1.1}^{+2.0}$ \\
$R_{\rm Refl}$ & $1.1_{-0.3}^{+0.5}$ & $0.8_{-0.2}^{+0.4}$ & $0.7_{-0.2}^{+0.6}$ & $0.8_{-0.5}^{+0.2}$ & $1.2_{-0.3}^{+0.3}$ & $1.0_{-0.2}^{+0.4}$ & $0.8_{-0.2}^{+0.6}$ & $0.6_{-0.2}^{+0.2}$ \\
$\theta$/$^\circ$ &-&-&-& $<17$&-&-&-& $<19$ \\
$C_{\rm FPMB/FPMA}$ & $1.06_{-0.018}^{+0.018}$ & $1.06_{-0.018}^{+0.018}$ & $1.06_{-0.018}^{+0.018}$ & $1.06_{-0.018}^{+0.018}$ & $1.03_{-0.015}^{+0.015}$ & $1.03_{-0.015}^{+0.015}$ & $1.03_{-0.015}^{+0.015}$ & $1.03_{-0.015}^{+0.015}$ \\
\hline
\multirow{2}{1cm}{$\chi^2$/d.o.f.} & 634/645 & 635/645 & 636/645 & 638/642 & 768/778 & 775/778 & 759/778 & 743/775\\
 & 0.98 & 0.98 & 0.99 & 0.99 & 0.98 & 1.00 & 0.98 & 0.96\\
\hline
\end{tabular}
\caption{Parameters of fits to \eso. Models are labelled by their primary component; each model also contains intrinsic absorption (with column density $N_{\rm H}$) and a cross-calibration constant between detectors ($C_{\rm FPMB/FPMA}$).}
\label{tab:esopars}
\end{table*}

\begin{table*}
\begin{tabular}{lcccccccc}
\hline
Dataset & \multicolumn{4}{c}{EGS} & \multicolumn{4}{c}{Cycle~3} \\
\cmidrule(lr){2-5}\cmidrule(lr){6-9}
Model & \textsc{pexmon} & \textsc{xillver} & \textsc{xillverCp} &  & \textsc{pexmon} & \textsc{xillver} & \textsc{xillverCp} &   \\
\hline
$N_{\rm H}/10^{22}$cm$^{-2}$ & $1.45_{-0.15}^{+0.15}$ & $1.45_{-0.15}^{+0.15}$ & $1.8_{-0.2}^{+0.2}$ && $1.2_{-0.2}^{+0.2}$ & $1.2_{-0.2}^{+0.2}$ & $1.8_{-0.2}^{+0.2}$  \\
$\Gamma$ & $1.53_{-0.03}^{+0.03}$ & $1.52_{-0.03}^{+0.03}$ & $1.72_{-0.01}^{+0.01}$ && $1.59_{-0.02}^{+0.01}$ & $1.59_{-0.02}^{+0.02}$ & $1.76_{-0.01}^{+0.01}$ \\
$E_{\rm Cut}$/keV & $78_{-12}^{+16}$ & $73_{-10}^{+13}$ & - && $80_{-9}^{+11}$ & $82_{-9}^{+12}$ & - & \\
$kT_{\rm e}$/keV & - & - & $19_{-2}^{+3}$ && - & - & $20_{-2}^{+3}$ \\
$A_{\rm Fe}$ & $>12$ & >6.5 & >7.9 && $>12$ & $10_{-4.5}^{+0}$ & $>8.3$ \\
$R_{\rm Refl}$ & $0.06_{-0.02}^{+0.02}$ & $0.065_{-0.025}^{+0.025}$ & $0.05_{-0.025}^{+0.025}$ && $0.25_{-0.05}^{+0.06}$ & $0.07_{-0.02}^{+0.02}$ & $0.06_{-0.02}^{+0.02}$\\
$C_{\rm FPMB/FPMA}$ & $1.02_{-0.01}^{+0.01}$ & $1.02_{-0.01}^{+0.01}$ & $1.02_{-0.01}^{+0.01}$ && $1.02_{-0.01}^{+0.01}$ & $1.02_{-0.01}^{+0.01}$ & $1.02_{-0.01}^{+0.01}$ \\
$C_{\rm XRT/FPMA}$ & $0.83_{-0.03}^{+0.03}$ & $0.83_{-0.03}^{+0.03}$ & $0.83_{-0.03}^{+0.03}$ && $0.83_{-0.04}^{+0.05}$ & $0.83_{-0.04}^{+0.05}$ & $0.86_{-0.04}^{+0.05}$ \\
\hline
\multirow{2}{1cm}{$\chi^2$/d.o.f.} & 1007/961 & 1010/961 & 1067/961 &   & 1290/1189 & 1291/1189 & 1340/1189 &  \\
& 1.05 &1.05&1.11&  & 1.09 & 1.09 & 1.13 &   \\
\hline
\end{tabular}
\caption{Parameters of fits to \igr. Models are labelled by their primary emission component; each model also contains Galactic absorption (with $N_{\rm H}=10^{22}$\,cm$^{-2}$) intrinsic absorption (with column density $N_{\rm H}$) and cross-calibration constants between detectors ($C_{\rm FPMB/FPMA}$, $C_{\rm XRT/FPMA}$).}
\label{tab:igrpars}
\end{table*}

\subsection{Spectral fitting}

We begin with a model with components to account for all of the spectral features mentioned. We use (\textsc{z})\textsc{tbabs} \citep{wilms00} for Galactic ($z=0$) and intrinsic (matched to source redshift) absorption. We do not include the Galactic component for \eso\ since this is insignificant compared to the intrinsic absorption.
We initially use \textsc{pexmon} to model the direct and reflected emission. This allows for a cut-off in direct coronal emission (modelled by an exponential cut-off) and reflection from neutral material with an iron-K$\alpha$ line, calculated self-consistently for a given metallicity. We allow the coronal parameters ($\Gamma$ and $E_{\rm Cut}$), reflection fraction and iron abundance to vary but freeze the inclination to the default value ($\theta=60^\circ$).

This provides reasonable fits to each dataset (Tables~\ref{tab:esopars},\ref{tab:igrpars}). 
The iron abundance for \igr\ is high ($A_{\rm Fe}>12$), although such high abundances have been found in other AGN \citep[e.g.][]{fabian09,ponti10}.
This could occur if there is significant enrichment of the nuclear gas by earlier generations of stars, through for example supernovae and stellar winds.

The cut-off energies, $130_{-60}^{+450}$ and $100_{-30}^{+90}$\,keV for \eso\ and $78_{-12}^{+16}$ and $80_{-9}^{+11}$\,keV for \igr, are consistent between observations for both sources and in agreement with at least some previous observations (\eso: $57_{-14}^{+18}$\,keV, \citealt{ricci17}; \igr: $79_{-15}^{+23}$\,keV, \citealt{molina07}).

The powerlaw indices are all relatively hard. \igr\ in particular has a very hard spectrum ($\Gamma=1.53\pm0.03$ and $1.52\pm0.03$) but not harder than has been found previously for this source \citep[$\Gamma=1.5$,][]{molina07}.

The sources differ markedly in their reflection fractions. While \eso\ has a reflection fraction around 1, as expected from illumination of a disc by an isotropic source away from strong relativistic effects, \igr\ has much weaker reflection ($R_{\rm Refl}=0.06\pm0.02$ and $0.25_{-0.05}^{+0.06}$).
Since \igr\ has a jet, this would fit with a scenario in which coronal material in \igr\ is the outflowing base of this jet and hence beamed away from the disc. Such a model has been proposed to explain the variability of Mrk~335 \citep{wilkins15} and the relationship between radio Eddington luminosity and X-ray reflection fraction \citep{king17}.

The cross-calibration between \nustar\ and \swift-XRT is slightly below that expected from IACHEC calibration observations \citep{madsen17} but not unreasonable when allowing for source variability.

\begin{figure*}
\centering
\includegraphics[width=\columnwidth]{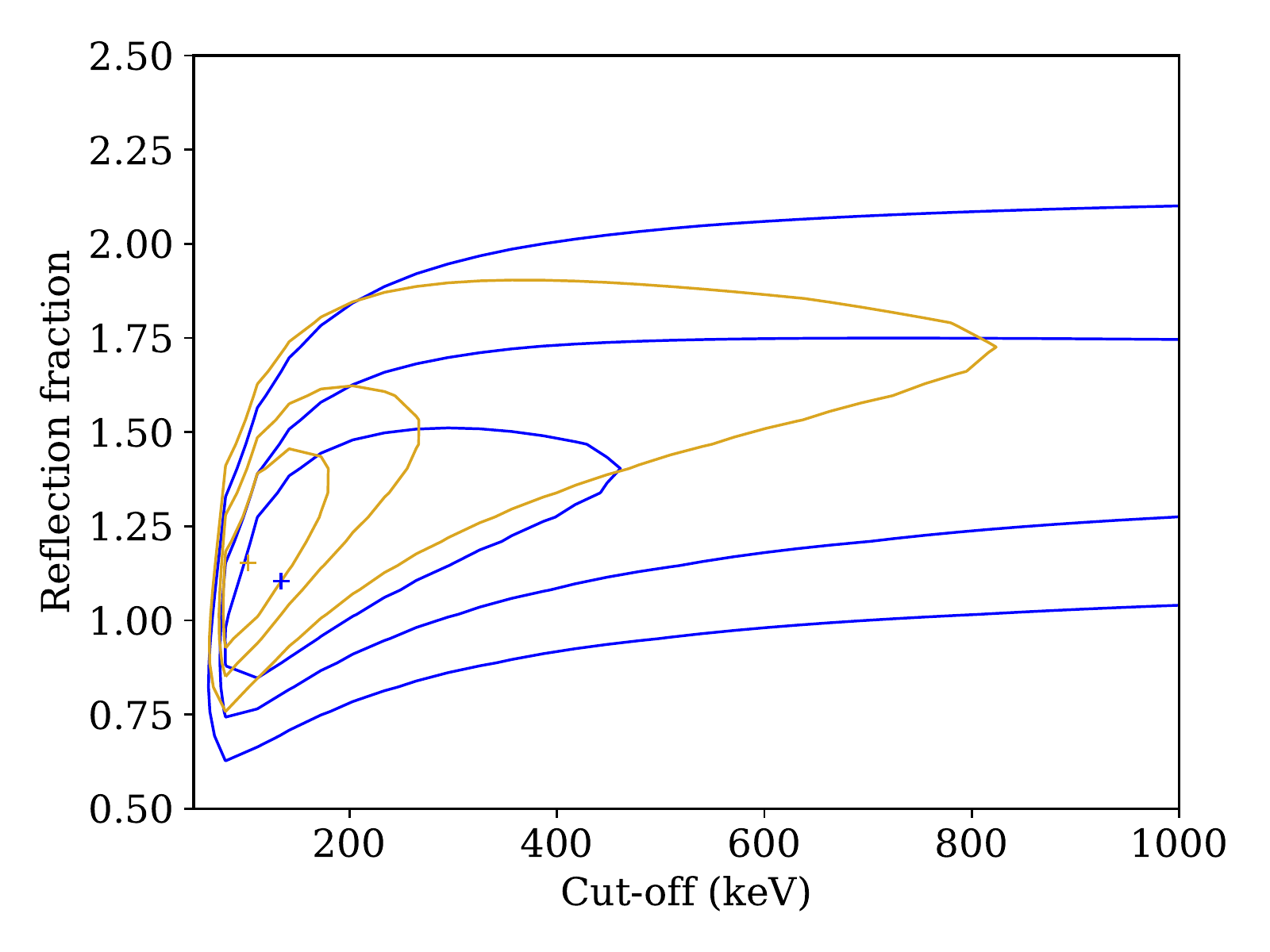}
\includegraphics[width=\columnwidth]{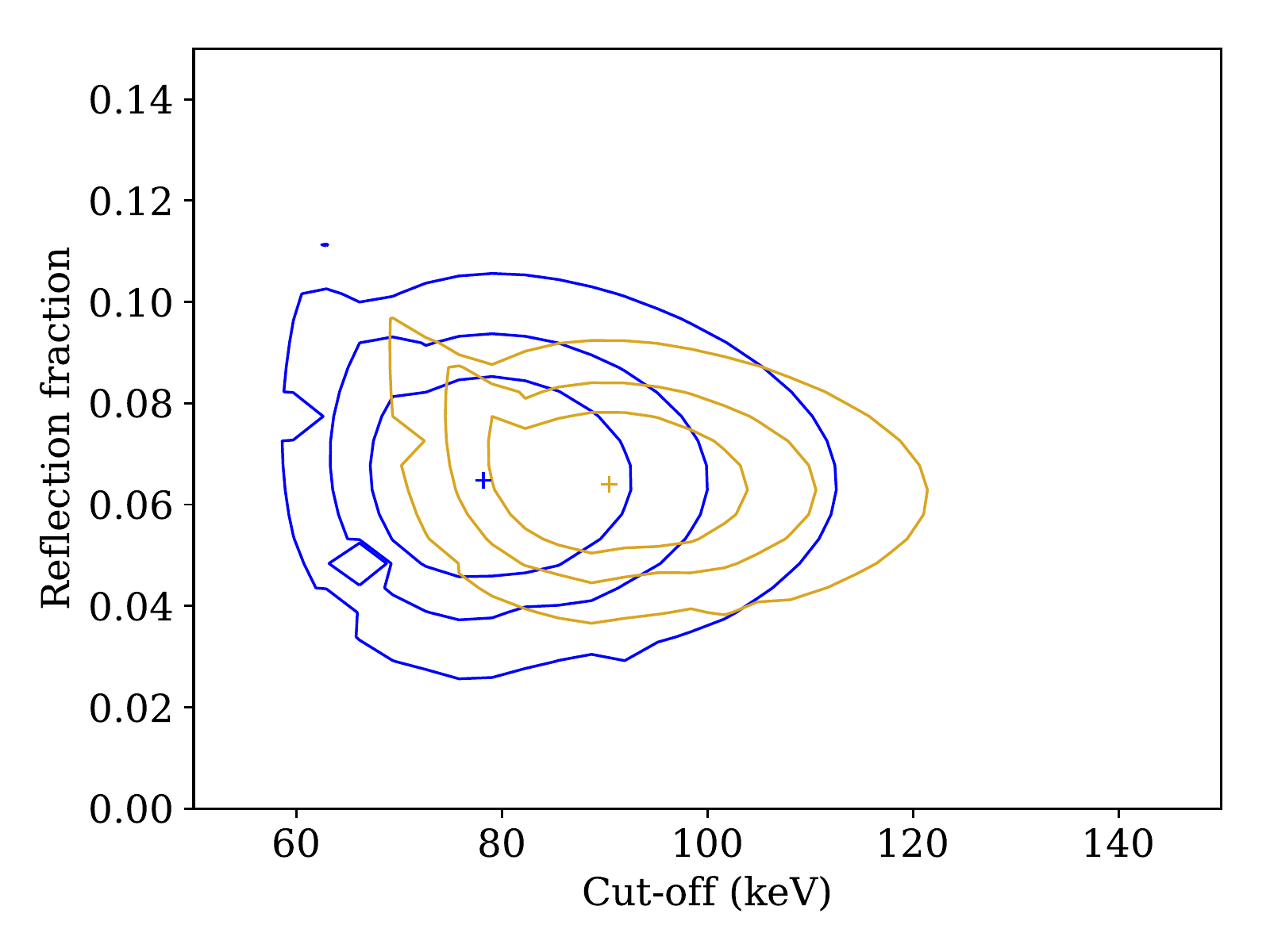}
\caption{
Contour plots of cut-off energy against reflection fraction for \eso\ (left) and \igr\ (right). The best fit is indicated by a cross, contours are shown at $1\sigma$, 90\% and $3\sigma$ confidence. Loci for EGS data are shown in blue, Cycle 3 in yellow. Despite some degeneracy between reflection strength and cut-off energy, both parameters are constrained.
}
\label{fig:contours}
\end{figure*}

There is inevitably some degeneracy between curvature due to the high-energy cut-off and due to reflection. To quantify this, we calculate confidence contours in the cut-off/reflection fraction plane (Figure~\ref{fig:contours}). This shows (particularly for \eso) the expected degeneracy, in that the fit has either a lower cut-off energy or a higher reflection fraction. However, in each case both parameters are still constrained (though only weakly for the shallowest, \eso\ EGS, observation).

To test the effect of different models, we perform a similar fit with the \textsc{xillver} model \citep{garcia13}, which has a more detailed model for the reflected spectrum. We fit for the same parameters as the \textsc{pexmon} model and fix the additional ionisation parameter $\log(\xi{\rm /erg\,cm\,s^{-1}})=0$ to best match the neutral \textsc{pexmon} reflection. This recovers very similar parameters (Tables~\ref{tab:esopars} and \ref{tab:igrpars}).

\subsubsection{Comptonisation models}

Having determined the shape of the high-energy roll-over phenomenologically, we now fit with physical Comptonisation models to obtain a direct constraint on the electron temperature.

We use the \textsc{xillverCp} model so that the reflected component is calculated self-consistently with the illuminating Comptonised continuum, which is generated with the \textsc{Nthcomp} model \citep{zdziarski96,zycki99}.
We again allow equivalent parameters to our previous models to be free. Fits to this model are given in Tables~\ref{tab:esopars},\ref{tab:igrpars}.
Most parameters are similar to those found for the previous models, but the fits to \igr\ have a significantly softer photon index ($\Gamma=1.72\pm0.01$ rather than $1.52\pm0.03$).

The electron temperatures for \eso\ are consistent with the expectation of a factor of $2-3$ lower than the cut-off energy \citep{petrucci01}. For \igr, this difference is slightly larger (around a factor of 4, though we note that the fit quality for \igr\ is not perfect). This could be due to the difference in shape of the reflected component (but this would be expected to have a larger effect in \eso, which has stronger reflection) or because the difference between $E_{\rm Cut}$ and $kT_{\rm e}$ becomes larger at high optical depth, which corresponds to a harder spectrum \citep{petrucci01}.

\subsubsection{Alternative models}

While the fit for \eso\ is formally acceptable, residuals are apparent around the iron line. Therefore, we also test a model with relativistically blurred reflection, using \textsc{relxillCp} \citep{dauser10,garcia14}. For the Cycle 3 observation, this gives a somewhat better fit, $\Delta\chi^2=15$ and shows only weak blurring ($R_{\rm in}>7\,R_{\rm ISCO}$). For the EGS observation, there is minimal improvement and parameters are consistent with the least blurring available to the model. Parameters of the Comptonised continuum are consistent with the unblurred model.
For completeness, we also fit this model to the observations of \igr\ but this does not provide a significant improvement.

We also consider a jet component in \igr: while \citet{tazaki10} estimate the contribution of a jet component to be subdominant, it is possible that even a small contribution has an effect on the more sensitive \nustar\ spectra presented here or that the jet emission has increased to a more significant level. Therefore, we also consider a model including a jet component approximated by a hard ($\Gamma<1.5$) power law. This reduces the best-fit value of the coronal temperature, as the high-energy coronal emission is replaced by the jet; the exact value depends on the index assumed for the jet component. If allowing any value of jet power, our coronal temperature measurement could then be seen as an upper limit. However, a strong jet component requires a $>78$\,keV flux far above the \swift-BAT value so would require a highly variable jet.
We therefore note this possible effect of jet emission but do not pursue the quantitative effect further.

\subsection{Comparison to other sources}

\begin{figure}
\includegraphics[width=\columnwidth]{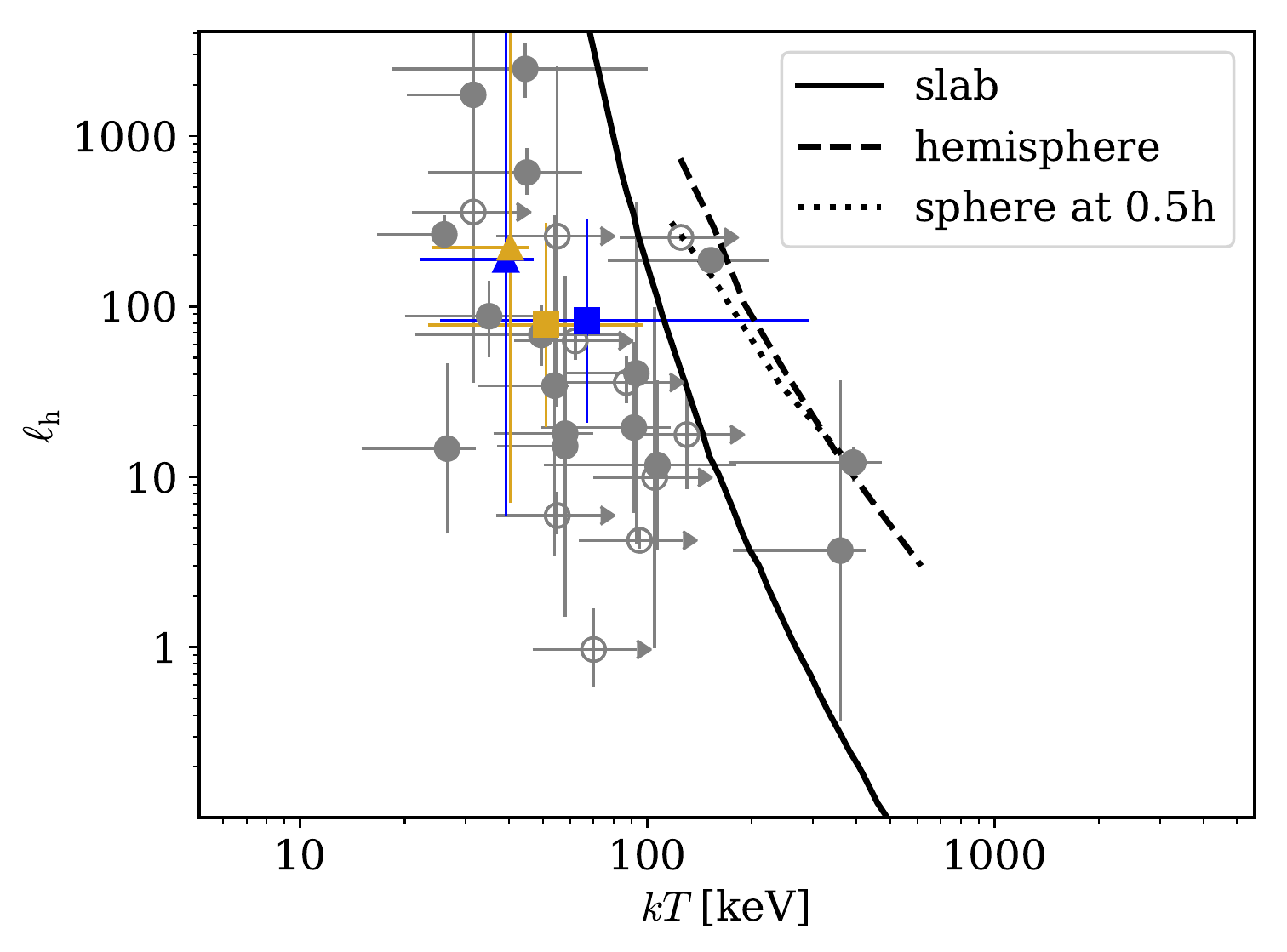}
\caption{Plot of coronal compactness ($\ell$) against temperature ($T$). Sources from \citet{fabian15} are shown by grey circles, \igr\ by triangles and \eso\ by squares. For \eso\ and \igr, EGS data is shown in blue and Cycle~3 in yellow. The limits due to pair production in various geometries are shown by the lines described in the legend.}
\label{fig:ellt}
\end{figure}

We compare the temperature and compactness of the coronae of \eso\ and \igr\ with that found for other sources by \citet{fabian15}. Using the formulae in \citet{fabian15}, we calculate compactness, $\ell$, and electron temperature, $kT_{\rm e}$ for each observation.
We take the required values of coronal luminosity and high-energy cut-off from the \textsc{pexmon} fit, since this is the most commonly used model in fits to the other sources in the sample.
Using values from the other models gives similar results.
Since we have no strong constraint on the coronal size, we follow \citet{fabian15} in using a fiducial value of $10r_{\rm g}$.
For \eso\ we use the mass estimate of \citet{czerny01},  $M_{\rm BH}=10^{7.1\pm0.6}\,{\rm M_{\odot}}$, and $M_{\rm BH}=10^{7.5\pm1.5}\,{\rm M_{\odot}}$ 
This constraint in the $\ell-T$ plane is shown in Figure~\ref{fig:ellt}.
Both sources have temperatures below the limit imposed by the pair thermostat and within the typical range of other sources of similar compactness.
The upper limits for \igr\ are significantly below the pair thermostat limit; this may indicate that some of the electrons in the corona have a non-thermal energy distribution \citep{fabian17}, as might the better description by an exponential roll-over than a thermal Comptonisation model.

\section{Discussion}
\label{section_discussion}

We have presented new hard X-ray spectra of two AGN made by \nustar\ and compared the coronal parameters found with the predictions of the pair thermostat for coronal temperature regulation.

Both sources have features which differ from the simplest typical AGN, which often formed the basis for the first round of \nustar\ observations. \eso\ has strong and variable obscuration ($\sim 1.7\pm0.2\times20^{23}$\,cm$^{-2}$ here, previously $1.0-1.7\times10^{23}$\,cm$^{-2}$, \citealt{warwick88}) and \igr\ has both a very hard spectrum ($\Gamma\sim1.5$) and significant radio emission \citep{ribo04,combi05}.
The strong obscuration makes measuring other spectral properties harder as their effects must be separated from features of obscuration. Since \nustar\ has good sensitivity up to high energies, we can still constrain features including the high-energy cut-off (which principally affects the spectrum at higher energies than obscuration) although to a lesser extent than might be possible with similarly deep observations of unobscured sources.
Despite their idiosyncrasies, both sources show coronal temperatures within the typical range for AGN (see Figure~\ref{fig:ellt}). This could indicate a controlled means of temperature regulation independent of the wider AGN environment, such as the pair thermostat discussed here. The apparent normality of sources which are in other ways unusual also provides a wider pool of AGN of which to take future coronal measurements.

Significant results on coronal properties have been based on large samples of low signal-to-noise spectra made with non-focussing instruments such as \integral\ and \swift-BAT.
The more sensitive \nustar\ spectra now available present an opportunity to cross-check results from previous instruments. The coronal temperature  of \eso\ agrees with that found from \swift-BAT \citep{ricci17} and that of \igr\ agrees with \integral\ \citep{malizia14}. This is promising for the robustness of results such as the decrease of cut-off energy with increasing Eddington rate \citep{ricci18} derived from such spectra.

We have also considered possible means of temperature regulation and found that both sources lie in the region of the $\ell-T$ plane allowed by the pair thermostat. The position relative to the annihilation limit is consistent with pair annihilation being an important means of regulation of the coronal temperature.
Furthermore, \igr\ has a temperature significantly below that implied by the pair thermostat. This could be due to the electron population including a non-thermal component, which tends to lower the limiting temperature \citep{fabian17}.

It is also possible that the compactness presented here is an under-estimate. Firstly, the $10r_{\rm g}$ size is a relatively high value: AGN coronae have often been found to be significantly smaller \citep[e.g.][]{parker14}, although this is usually accompanied by strong relativistic reflection. 
Secondly, the corona may have a highly inhomogeneous flux-density: it may composed of many much smaller regions of higher compactness within the overall $\sim10r_{\rm g}$ extent.
Both these effects would move the points upwards, closer to the pair-production limit.

\section*{Acknowledgements}

DJKB acknowledges financial support from the Science and Technology Facilities Council (STFC).
ACF acknowledges support from the ERC Advanced Grant FEEDBACK 340442.
This work made use of data from the \nustar\ mission, a project led by the California Institute of Technology, managed by the Jet Propulsion Laboratory, and funded by the National Aeronautics and Space Administration. This research has made use of the \nustar\ Data Analysis Software (NuSTARDAS) jointly developed by the ASI Science Data Center (ASDC, Italy) and the California Institute of Technology (USA).
This work made use of data supplied by the UK Swift Science Data Centre at the University of Leicester.

\bibliographystyle{mnras}
\bibliography{paper}

\bsp	
\label{lastpage}
\end{document}